# Eigenvalue asymptotics of the even-dimensional exterior Landau-Neumann Hamiltonian

Mikael Persson

**Abstract**: We study the Schrödinger operator with a constant magnetic field in the exterior of a compact domain in $\mathbb{R}^{2d}$, $d \geq 1$. The spectrum of this operator consists of clusters of eigenvalues around the Landau levels. We give asymptotic formulas for the rate of accumulation of eigenvalues in these clusters. When the compact is a Reinhart domain we are able to show a more precise asymptotic formula.

## 1 Introduction

The Landau Hamiltonian describes a charged particle moving in a plane, influenced by a constant magnetic field of strength $B > 0$ orthogonal to the plane. It is a classical result, see [Foc28, Lan30], that the spectrum of the Landau Hamiltonian consists of infinitely degenerate eigenvalues $B(2q + 1)$, $q = 0, 1, 2, \ldots$, called Landau levels.

In this paper we will study the even-dimensional Landau Hamiltonian outside a compact obstacle, imposing magnetic Neumann conditions at the boundary. Our motivation to study this operator comes mainly from the papers [HS02, PR07]. Spectral properties of the exterior Landau Hamiltonian in the plane are discussed in [HS02], under both Dirichlet and Neumann conditions at the boundary, with focus mainly on properties of the eigenfunctions. A more qualitative study of the spectrum is done in [PR07], where the authors fix an interval around a Landau level and describe how fast the eigenvalues in that cluster converges to that Landau level. They work in the plane and with Dirichlet boundary conditions only. The goal of this paper is to perform the same qualitative description when we impose magnetic Neumann conditions at the boundary. Moreover, we do not limit ourself to the plane, but work in arbitrary evendimensional Euclidean space.

The result is that the eigenvalues do accumulate with the same rate to the Landau levels for both types of boundary conditions, see Theorem 3.2 for the details. However, the eigenvalues can only accumulate to a Landau level from below in the Neumann setting. In the Dirichlet case they accumulate only from above.

It should be mentioned that we suppose that the compact set removed has no holes and that its boundary is smooth. This is far more restrictive than the conditions imposed on the compact set in [PR07].

Several different perturbations of the Landau Hamiltonian have been studied in the last years, see [Raĭ90, RW02a, RW02b, MR03, Rai03, FP06, RS08, PR07]. They all share the common idea of making a reduction to a certain Toeplitz-type operator whose spectral asymptotics is known. We also do this kind of reduction. The method we use is based on the theory for pseudodifferential operators and boundary PDE methods, which we have not seen in any of the mentioned papers.

In Section 2 we define the Landau Hamiltonian and give some auxiliary results about its spectrum, eigenspaces and Green function.

We begin Section 3 by defining the exterior Landau Hamiltonian with magnetic Neumann boundary condition and formulate and prove the main theorems (Theorem 3.1 and 3.2) about the spectral asymptotics of the operator. The main part of the proof, the reduction step, is quite technical and therefore moved to Section 4. When the reduction step is done we use the asymptotic formulas of the spectrum of the Toeplitz-type operators, given in [FP06, MR03], to obtain the asymptotic formulas in Theorem 3.2.

In the higher dimensional case ($\mathbb{R}^{2d}$, $d > 1$) we also consider the case when the compact obstacle is a Reinhart domain. We use some ideas from [Par94] to prove a more precise asymptotic formula for the eigenvalues. This is done in Section 5.

## 2  The Landau Hamiltonian in $\mathbb{R}^{2d}$

We denote by $x = (x^1, \ldots, x^{2d})$ a point in $\mathbb{R}^{2d}$. Let $B > 0$ and denote by $\vec{a}$ the magnetic vector potential

$$\vec{a}(x) = \big(a_1(x), \ldots, a_{2d}(x)\big) = \frac{B}{2}\left(-x^2, x^1, -x^4, x^3, \ldots, -x^{2d}, x^{2d-1}\right).$$

It corresponds to an isotropic magnetic field of constant strength $B$. The Landau Hamiltonian $L$ in $\mathbb{R}^{2d}$ describes a charged, spinless particle in this homogeneous magnetic field. It is given by

$$L = (-i\nabla - \vec{a})^2,$$

and is essentially self-adjoint on the set $C_0^\infty(\mathbb{R}^{2d})$ in the usual Hilbert space $\mathcal{H} = L_2(\mathbb{R}^{2d})$. For $j = 1, \ldots, d$ we also introduce the self-adjoint operators

$$L_j = \left(-i\left(\frac{\partial}{\partial x^{2j-1}}, \frac{\partial}{\partial x^{2j}}\right) - (a_{2j-1}, a_{2j})\right)^2,$$

in the Hilbert spaces $\mathcal{H}_j = L_2(\mathbb{R}^2)$. Note that $\mathcal{H} = \bigotimes_{j=1}^d \mathcal{H}_j$, and

$$L = L_1 \otimes I^{\otimes(d-1)} + I \otimes L_2 \otimes I^{\otimes(d-2)} + \ldots + I^{\otimes(d-1)} \otimes L_d. \tag{2.1}$$

## 2.1 Landau levels

The spectrum of each twodimensional Landau Hamiltonian $L_j$ consists of so called Landau levels, eigenvalues $B(2q+1)$, $q \in \mathbb{N}$, each of infinite multiplicity. Let $\hat{\kappa} = (\kappa_1, \ldots, \kappa_d) \in \mathbb{N}^d$ be a multiindex. We denote by $|\hat{\kappa}| = \kappa_1 + \ldots + \kappa_d$ the length of the multiindex $\hat{\kappa}$ and also set $\hat{\kappa}! = \kappa_1! \cdot \ldots \cdot \kappa_d!$. From (2.1) it follows that the spectrum of $L$ consists of the infinitely degenerate eigenvalues

$$\Lambda_{\hat{\kappa}} = B \sum_{j=1}^{d}(2\kappa_j + 1), \quad \kappa_j \in \mathbb{N}.$$

Note that $\Lambda_{\hat{\kappa}} = \Lambda_{\hat{\kappa}'}$ if $|\hat{\kappa}| = |\hat{\kappa}'|$. Hence the spectrum of $L$ consists of eigenvalues of the form $\Lambda_\mu = B(2\mu + d)$, $\mu \in \mathbb{N}$.

## 2.2 Creation and annihilation operators

The structure of the eigenspaces of $L$ has been described before in [MR03]. We give the results without proofs. It is convenient to introduce complex notation. Let $z = (z^1, \ldots, z^d) \in \mathbb{C}^d$, where $z^j = x^{2j-1} + ix^{2j}$. Also, we use the scalar potential $W(z) = -\frac{B}{4}|z|^2$ and the complex derivatives

$$\frac{\partial}{\partial z^j} = \frac{1}{2}\left(\frac{\partial}{\partial x^{2j-1}} - i\frac{\partial}{\partial x^{2j}}\right), \quad \frac{\partial}{\partial \bar{z}^j} = \frac{1}{2}\left(\frac{\partial}{\partial x^{2j-1}} + i\frac{\partial}{\partial x^{2j}}\right).$$

We define creation and annihilation operators $\mathfrak{Q}_j^*$, $\mathfrak{Q}_j$ as

$$\mathfrak{Q}_j^* = -2ie^{-W}\frac{\partial}{\partial z^j}e^{W}, \qquad \mathfrak{Q}_j = -2ie^{W}\frac{\partial}{\partial \bar{z}^j}e^{-W},$$

and note that

$$\left[\mathfrak{Q}_j^*, \mathfrak{Q}_k^*\right] = \left[\mathfrak{Q}_j, \mathfrak{Q}_k\right] = \left[\mathfrak{Q}_j^*, \mathfrak{Q}_k\right] = 0, \quad \text{if } j \neq k. \tag{2.2}$$

The notation $\mathfrak{Q}_j^*$ for the creation operators is motivated by the fact that it is the formal adjoint of $\mathfrak{Q}_j$ in $\mathcal{H}$.

A function $u$ belongs to the lowest Landau level $\Lambda_0$ if and only if $\mathfrak{Q}_j u = 0$ for $j = 1, \ldots, d$. This means that the function $f = e^{-W}u$ is an entire function, so via multiplication by $e^{-W}$ the eigenspace $\mathfrak{L}_{\Lambda_0}$ corresponding to $\Lambda_0$ is equivalent to the Fock space

$$\mathfrak{F}_B^2 = \left\{ f \mid f \text{ is entire and } \int_{\mathbb{C}^d} |f|^2 e^{-\frac{B}{2}|z|^2} \, dm(z) < \infty \right\}.$$

Here, and elsewhere, $dm$ denotes the Lebesgue measure. A function $u$ belongs to the eigenspace $\mathfrak{L}_{\Lambda_\mu}$ of the Landau level $\Lambda_\mu$ if and only if it can be written in the form

$$u = \sum_{|\hat{\kappa}|=\mu} c_{\hat{\kappa}} (\mathfrak{Q}^*)^{\hat{\kappa}} (e^W f_{\hat{\kappa}}),$$

where $(\mathfrak{Q}^*)^{\hat{\kappa}} = (\mathfrak{Q}_1^*)^{\kappa_1} \cdots (\mathfrak{Q}_d^*)^{\kappa_d}$ and $f_{\hat{\kappa}}$ all belong to $\mathfrak{F}_B^2$. The multiplicity of the eigenvalue $\Lambda_\mu$ is equal to $\binom{\mu+d-1}{d-1}$. We denote by $\mathcal{P}_{\Lambda_{\hat{\kappa}}}$ and $\mathcal{P}_{\Lambda_\mu}$ the projection onto the eigenspaces $\mathfrak{L}_{\Lambda_{\hat{\kappa}}}$ and $\mathfrak{L}_{\Lambda_\mu}$ respectively, and note by (2.2) that the orthogonal decompositions

$$\mathfrak{L}_{\Lambda_\mu} = \bigoplus_{|\hat{\kappa}|=\mu} \mathfrak{L}_{\Lambda_{\hat{\kappa}}}, \quad \mathcal{P}_{\Lambda_\mu} = \bigoplus_{|\hat{\kappa}|=\mu} \mathcal{P}_{\Lambda_{\hat{\kappa}}} \tag{2.3}$$

hold in $\mathcal{H}$.

### 2.3 The resolvent

Let $R_\rho = (L + \rho I)^{-1}$ be the resolvent of $L$, $\rho \geq 0$. An explicit formula of the kernel $G_\rho(x, y)$ of $R_\rho$ was given in [HS02] for $d = 1$. In Section 4.2 we will use the behavior of $G_\rho(x, y)$ near the diagonal $x = y$, given in the following lemma.

**Lemma 2.1** *$R_\rho$ is an integral operator with kernel $G_\rho(x, y)$ that has the following singularity at the diagonal,*

$$G_\rho(x, y) \sim \begin{cases} \frac{1}{\pi} \log(1/|x-y|) + O(1), & d = 1; \\ \frac{1}{2\pi^2} |x-y|^{-2} + O(\log(1/|x-y|)), & d = 2; \\ \frac{\Gamma(d-1)}{2\pi^d} |x-y|^{2-2d} + O(|x-y|^{4-2d}), & d > 2; \end{cases} \tag{2.4}$$

*as $|x - y| \to 0$.*

**Proof** The kernel $G_\rho(x, y)$ of $R_\rho$ can be written as

$$G_\rho(x, y) = \int_0^\infty e^{-\rho t} e^{-Lt}(x, y) \, dm(t).$$

Now, since the variables separate pairwise, we have

$$e^{-Lt}(x, y) = \prod_{j=1}^d e^{-L_j t}(x^{2j-1}, x^{2j}, y^{2j-1}, y^{2j}).$$

The formula for $e^{-L_j t}$ is given in [Sim79a]. It reads

$$e^{-L_j t} = \frac{B}{4\pi} \exp\left(-\frac{iB}{2}\left(x^{2j-1}y^{2j} - x^{2j}y^{2j-1}\right)\right)\frac{1}{\sinh(Bt/2)} \times$$

$$\times \exp\left(-\frac{B}{4}\coth(Bt/2)\left[\left(x^{2j-1} - y^{2j-1}\right)^2 + \left(x^{2j} - y^{2j}\right)^2\right]\right)$$

Hence the formula for $G_\rho(x,y)$ becomes

$$G_\rho(x,y) = \left(\frac{B}{4\pi}\right)^d \exp\left(-\frac{iB}{2}\sum_{j=1}^d (x^{2j-1}y^{2j} - x^{2j}y^{2j-1})\right) I(|x-y|^2) \tag{2.5}$$

where

$$I(s) = \int_0^\infty e^{-\rho t} \frac{1}{\sinh^d(Bt/2)} \exp\left(-\frac{B}{4}\coth(Bt/2)s\right) \, dm(t).$$

An expansion of $I(s)$ shows that

$$I(s) \sim \begin{cases} \left(\frac{2}{B}\right)\log(1/s) + O(1), & d = 1; \\ \frac{8}{B^2}s^{-1} + O(\log(1/s)), & d = 2; \\ \left(\frac{4}{B}\right)^d \frac{\Gamma(d-1)}{2} s^{1-d} + O(s^{2-d}), & d > 2; \end{cases} \quad \text{as } s \to 0,$$

from which (2.4) follows. $\square$

## 3 The exterior Landau-Neumann Hamiltonian in $\mathbb{R}^{2d}$

Let $K \subset \mathbb{R}^{2d}$ be a simply connected compact domain with smooth boundary $\Gamma$ and let $\Omega = \mathbb{R}^{2d} \setminus K$. We define the exterior Landau-Neumann Hamiltonian $L_\Omega$ in $\mathcal{H}_\Omega = L_2(\Omega)$ by

$$L_\Omega = (-i\nabla - \vec{a})^2, \quad \text{in } \Omega; \tag{3.1}$$

with magnetic Neumann boundary conditions

$$\partial_N u := (-i\nabla - \vec{a})u \cdot \nu = 0 \quad \text{on } \Gamma. \tag{3.2}$$

Here $\nu$ denotes the exterior normal to $\Gamma$. Our aim is to study how much the spectrum of $L_\Omega$ differs from the Landau levels discussed in the previous section. The first Theorem below states that the eigenvalues of $L_\Omega$ can accumulate to each Landau level only from below. The second Theorem says that the eigenvalues do accumulate to the Landau levels from below, and the rate of convergence is given.

**Theorem 3.1** *For every $\mu \in \mathbb{N}$ and each $\varepsilon$, $0 < \varepsilon < dB$, the number of eigenvalues of $L_\Omega$ in the interval $(\Lambda_\mu, \Lambda_\mu + \varepsilon)$ is finite.*

Denote by $l_1^{(\mu)} \leq l_2^{(\mu)} \leq \cdots$ the eigenvalues of $L_\Omega$ in the interval $(\Lambda_{\mu-1}, \Lambda_\mu)$ and $N(a, b, T)$ the number of eigenvalues of the operator $T$ in the interval $(a, b)$, counting multiplicities. Also, let $\text{Cap}(K)$ denote the logarithmic capacity of $K$, see [Lan72].

**Theorem 3.2** *Let $\mu \in \mathbb{N}$.*

(a) *If $d = 1$ then $\lim_{j \to \infty} \left( j! \left( \Lambda_\mu - l_j^{(\mu)} \right) \right)^{1/j} = \frac{B}{2} \left( \text{Cap}(K) \right)^2$.*

(b) *If $d > 1$ then $N(\Lambda_{\mu-1}, \Lambda_\mu - \lambda, L_\Omega) \sim \binom{\mu+d-1}{d-1} \frac{1}{d!} \left( \frac{|\ln \lambda|}{\ln |\ln \lambda|} \right)^d$ as $\lambda \searrow 0$.*

### 3.1 Proof of the Theorems

We want to compare the spectrum of the operators $L$ and $L_\Omega$. However, the expression $L - L_\Omega$ has no meaning since $L$ and $L_\Omega$ acts in different Hilbert spaces. We introduce the Hilbert space $\mathcal{H}_K = L_2(K)$ and define the interior Landau-Neumann Hamiltonian $L_K$ in $\mathcal{H}_K$ by the same formulas as in (3.1) and (3.2) but with $\Omega$ replaced by $K$. We note that $\mathcal{H} = \mathcal{H}_K \oplus \mathcal{H}_\Omega$ and define $\tilde{L}$ as

$$\tilde{L} = L_K \oplus L_\Omega, \quad \text{in } \mathcal{H}_K \oplus \mathcal{H}_\Omega.$$

The inverse of $L_K$ is compact, so $L_K$ has at most a finite number of eigenvalues in each interval $(\Lambda_{\mu-1}, \Lambda_\mu)$. The operators $L_K$ and $L_\Omega$ act in orthogonal subspaces of $\mathcal{H}$, so $\sigma(\tilde{L}) = \sigma(L_K) \cup \sigma(L_\Omega)$. This means that $\tilde{L}$ has the same spectral asymptotics as $L_\Omega$ in each interval $(\Lambda_{\mu-1}, \Lambda_\mu)$, so it is enough to prove the statements in Theorem 3.1 and 3.2 for the operator $\tilde{L}$ instead of $L_\Omega$.

Since the unbounded operators $L$ and $\tilde{L}$ have different domains, we cannot compare them directly. However, they act in the same Hilbert space, so we can compare their inverses. Let

$$R = R_0 = L^{-1}, \quad \text{and} \quad \tilde{R} = \tilde{L}^{-1} = L_K^{-1} \oplus L_\Omega^{-1},$$

and set

$$V = \tilde{R} - R, \quad \text{and} \quad T_\mu = \mathcal{P}_{\Lambda_\mu} V \mathcal{P}_{\Lambda_\mu}, \quad \text{for } \mu \in \mathbb{N}.$$

**Lemma 3.3** *$V$ is non-negative and compact.*

**Proof** See Section 4.1. □

By Weyl's theorem the essential spectrum of $R$ and $\tilde{R}$ coincides. Since $\tilde{R} = R + V$ and $V \geq 0$, Theorem 3.1 follows immediately from Theorem 9.4.7 in [BS87] and the fact that $\sigma(R) = \sigma_{\text{ess}}(R) = \{\Lambda_\mu^{-1}\}$. We continue with the proof of Theorem 3.2.

Let $\tau > 0$ be such that $\left((\Lambda_\mu^{-1} - 2\tau, \Lambda_\mu^{-1} + 2\tau) \setminus \{\Lambda_\mu^{-1}\}\right) \cap \sigma_{\text{ess}}(R) = \emptyset$. Denote the eigenvalues of $T_\mu$ by

$$t_1^{(\mu)} \geq t_2^{(\mu)} \geq \cdots,$$

and the eigenvalues of $\tilde{R}$ in the interval $(\Lambda_\mu^{-1}, \Lambda_\mu^{-1} + \tau)$ by

$$r_1^{(\mu)} \geq r_2^{(\mu)} \geq \cdots.$$

**Lemma 3.4** *Given $\varepsilon > 0$ there exists an integer $l$ such that*

$$(1-\varepsilon)t_{j+l}^{(\mu)} \leq r_j^{(\mu)} - \Lambda_\mu^{-1} \leq (1+\varepsilon)t_{j-l}^{(\mu)}, \quad \text{for all sufficiently large } j.$$

**Proof** See Proposition 2.2 in [PR07]. □

Hence the study of the asymptotics of the eigenvalues of $\tilde{R}$ is reduced to the study of the eigenvalues of the Toeplitz-type operator $T_\mu$. For a bounded simply connected set $U$ in $\mathbb{R}^{2d}$ we define the Toeplitz operator $S_\mu^U$ as

$$S_\mu^U = \mathcal{P}_{\Lambda_\mu} \chi_U \mathcal{P}_{\Lambda_\mu},$$

where $\chi_U$ denotes the characteristic function of $U$. The following lemma reduces our problem to the study of these Toeplitz operators, which are easier to study than $T_\mu$.

**Lemma 3.5** *Let $K_0 \Subset K \Subset K_1$ be compact domains such that $\partial K_i \cap \Gamma = \emptyset$. There exist a constant $C > 0$ and a subspace $\mathcal{S} \subset \mathcal{H}$ of finite codimension such that*

$$\frac{1}{C}\langle f, S_\mu^{K_0} f\rangle \leq \langle f, T_\mu f\rangle \leq C\langle f, S_\mu^{K_1} f\rangle \tag{3.3}$$

*for all $f \in \mathcal{S}$.*

**Proof** See Section 4.2. □

The asymptotic expansion of the spectrum of $S_\mu^U$ is given in the following lemma.

**Lemma 3.6** *Denote by $s_1^{(\mu)} \geq s_2^{(\mu)} \geq \ldots$ the eigenvalues of $S_\mu^U$ and by $n(\lambda, S_\mu^U)$ the number of eigenvalues of $S_\mu^U$ greater than $\lambda$ (counting multiplicity). Then*

(a) *if $d = 1$ we have $\lim_{j \to \infty} \left(j! s_j^{(\mu)}\right)^{1/j} = \frac{B}{2}\left(\text{Cap}(U)\right)^2$,*

(b) *if $d > 1$ we have $n(\lambda, S_\mu^U) \sim \binom{\mu+d-1}{d-1}\frac{1}{d!}\left(\frac{|\ln \lambda|}{\ln|\ln \lambda|}\right)^d$ as $\lambda \searrow 0$.*

**Proof** See Lemma 3.2 in [FP06] for part (a) and Proposition 7.1 in [MR03] for part (b). □

We are now able to finish the proof of Theorem 3.2. By letting $K_0$ and $K_1$ in Lemma 3.5 get closer and closer to our compact $K$ we see that the eigenvalues $\{t_j^{(\mu)}\}$ of $T_\mu$ satisfy

$$\lim_{n\to\infty} \left(j! t_j^{(\mu)}\right)^{1/j} = \frac{B}{2}\left(\mathrm{Cap}(K)\right)^2 \tag{3.4}$$

if $d = 1$, and

$$n(\lambda, T_\mu) \sim \binom{\mu+d-1}{d-1}\frac{1}{d!}\left(\frac{|\ln\lambda|}{\ln|\ln\lambda|}\right)^d, \quad \text{as } \lambda \searrow 0 \tag{3.5}$$

if $d > 1$. Since neither of the formulas (3.4) nor (3.5) are sensitive for finite shifts in the indices it follows from Lemma 3.4 that the eigenvalues of $\{r_j^{(\mu)}\}$ $\tilde{R}$ satisfies

$$\lim_{j\to\infty} \left(j!(r_j^{(\mu)} - \Lambda_\mu^{-1})\right)^{1/j} = \frac{B}{2}\left(\mathrm{Cap}(K)\right)^2$$

if $d = 1$, and

$$N(\Lambda_\mu^{-1} + \lambda, \Lambda_{\mu-1}^{-1}, \tilde{R}) \sim \binom{\mu+d-1}{d-1}\frac{1}{d!}\left(\frac{|\ln\lambda|}{\ln|\ln\lambda|}\right)^d, \quad \text{as } \lambda \searrow 0$$

If we translate this in terms of $\tilde{L}$ we get

$$\lim_{j\to\infty} \left(j!(\Lambda_\mu - l_j^{(\mu)})\right)^{1/j} = \frac{B}{2}\left(\mathrm{Cap}(K)\right)^2$$

for $d = 1$, and

$$N(\Lambda_{\mu-1}, \Lambda_\mu - \lambda, \tilde{L}) \sim \binom{\mu+d-1}{d-1}\frac{1}{d!}\left(\frac{\left|\ln\frac{\lambda}{\Lambda_\mu(\Lambda_\mu-\lambda)}\right|}{\ln\left|\ln\frac{\lambda}{\Lambda_\mu(\Lambda_\mu-\lambda)}\right|}\right)^d$$

$$\sim \binom{\mu+d-1}{d-1}\frac{1}{d!}\left(\frac{|\ln\lambda|}{\ln|\ln\lambda|}\right)^d, \quad \text{as } \lambda \searrow 0,$$

for $d > 1$. This completes the proof of Theorem 3.2. □

## 4 Proof of the Lemmas

In this section we prove Lemma 3.3 and 3.5.

### 4.1 Proof of Lemma 3.3

The operators $L$ and $\tilde{L}$ are defined by the same expression, but the domain of $\tilde{L}$ is contained in the domain of $L$. It follows from Proposition 2.1 in [PR07] that $L - \tilde{L} \geq 0$, and hence $V = \tilde{R} - R \geq 0$.

Next we prove the compactness of $V$. Let $f$ and $g$ belong to $\mathcal{H}$. Also, let $u = Rf$ and $v = \tilde{R}g$. Then $u$ belongs to the domain of $L$ and $v$ belongs to the domain of $\tilde{L}$, so $v = v_K \oplus v_\Omega$, and $L_K v_K \oplus L_\Omega v_\Omega = g$. Integrating by parts and using (3.2) for $v_K$ and $v_\Omega$, we get

$$\begin{aligned}
\langle f, Vg \rangle &= \langle f, \tilde{R}g \rangle - \langle Rf, g \rangle \\
&= \int_K Lu \cdot \overline{v_K}\, dm(x) + \int_\Omega Lu \cdot \overline{v_\Omega}\, dm(x) \\
&\quad - \int_K u \cdot \overline{L_K v_K}\, dm(x) - \int_\Omega u \cdot \overline{L_\Omega v_\Omega}\, dm(x) \\
&= \int_\Gamma \partial_N u \cdot \overline{(v_\Omega - v_K)}\, dS.
\end{aligned} \quad (4.1)$$

Here $dS$ denotes the surface measure on $\Gamma$.

Take a smooth cut-off function $\chi \in C_0^\infty(\mathbb{R}^{2d})$ such that $\chi(x) = 1$ in a neighborhood of $K$. Then we can replace $u$ and $v$ by $\tilde{u} = \chi u$ and $\tilde{v} = \chi v$ in the right hand side of (4.1). By local elliptic regularity we have that $\tilde{u} \in H^2(\mathbb{R}^{2d})$ and $\tilde{v} \in H^2(\mathbb{R}^{2d} \setminus \Gamma)$. However, the operator $\tilde{u} \mapsto \partial_N \tilde{u}|_\Gamma$ is compact as considered from $H^2(\mathbb{R}^{2d})$ to $L_2(\Gamma)$ and both $\tilde{v} \mapsto \tilde{v}_\Omega|_\Gamma$ and $\tilde{v} \mapsto \tilde{v}_K|_\Gamma$ are compact as considered from $H^2(\mathbb{R}^{2d} \setminus \Gamma)$ to $L_2(\Gamma)$, so it follows that $V$ is compact. $\square$

### 4.2 Proof of Lemma 3.5

We start by showing that $T_\mu$ can be considered as an elliptic Pseudodifferential operator of order 1 on some subspace of $L_2(\Gamma)$ of finite codimension, and hence that there exists a constant $C > 0$ such that

$$\frac{1}{C}\|f\|_{L_2(\Gamma)}\|f\|_{H^1(\Gamma)} \leq \langle f, T_\mu f \rangle \leq C\|f\|_{L_2(\Gamma)}\|f\|_{H^1(\Gamma)} \quad (4.2)$$

for all $f$ in that subspace.

Let $f$ and $g$ belong to $\mathcal{H}$. Also, let $u = Rf$, $v = \tilde{R}g$ and $w = Rg$. We saw in (4.1) that

$$\langle f, Vg \rangle = \int_\Gamma \partial_N u \cdot \overline{(v_\Omega - v_K)}\, dS.$$

To go further we will introduce the Neumann to Dirichlet and Dirichlet to Neumann operators. Let $G_\rho(x, y)$ be as in (2.5). We start with the single and double layer integral operators, defined by

$$\mathcal{A}\alpha(x) = \int_\Gamma G_0(x - y)\alpha(y)\,dS(y), \quad x \in \mathbb{R}^{2d},$$

$$\mathcal{B}\alpha(x) = \int_\Gamma \partial_{N_y} G_0(x - y)\alpha(y)\,dS(y), \quad x \in \mathbb{R}^{2d} \setminus \Gamma,$$

$$A\alpha(x) = \int_\Gamma G_0(x - y)\alpha(y)\,dS(y), \quad x \in \Gamma, \text{ and}$$

$$B\alpha(x) = \int_\Gamma \partial_{N_y} G_0(x - y)\alpha(y)\,dS(y), \quad x \in \Gamma.$$

The last two operators are compact on $L_2(\Gamma)$, since, by Lemma 2.1, their kernels have weak singularities. Moreover, since the kernel $G_0$ has the same singularity as the Green kernel for the Laplace operator in $\mathbb{R}^{2d}$ (see [Tay96b]), we have the following limit relations on $\Gamma$

$$\mathcal{A}\alpha_K = A\alpha_K, \qquad \mathcal{B}\alpha_K = \frac{1}{2}\alpha + B\alpha,$$

$$\mathcal{A}\alpha_\Omega = A\alpha_\Omega, \qquad \mathcal{B}\alpha_\Omega = -\frac{1}{2}\alpha + B\alpha. \tag{4.3}$$

Using a Green-type formula for $L$ in $K$ we see that

$$\beta = \mathcal{B}\beta_K - \mathcal{A}(\partial_N \beta_K).$$

If we combine this with the limit relations (4.3) we get

$$\left(B - \frac{1}{2}I\right)\beta_K = A(\partial_N \beta_K), \quad \text{on } \Gamma.$$

A similar calculation for $\Omega$ gives

$$\left(B + \frac{1}{2}I\right)\beta_\Omega = A(\partial_N \beta_\Omega), \quad \text{on } \Gamma.$$

It seems natural to do the following definitions.

**Definition 4.1** We define the Dirichlet-to-Neumann and Neumann-to-Dirichlet operators in $K$ and $\Omega$ as

$$(DN)_K = A^{-1}\left(B - \frac{1}{2}I\right), \quad (ND)_K = \left(B - \frac{1}{2}I\right)^{-1} A,$$

$$(DN)_\Omega = A^{-1}\left(B + \frac{1}{2}I\right), \quad (ND)_\Omega = \left(B + \frac{1}{2}I\right)^{-1} A.$$

**Remark 4.1** The inverses above exist at least on a space of finite codimension. This follows from the fact that $A$ is elliptic and $B$ is compact.

**Lemma 4.1** *The operator $(ND)_K - (ND)_\Omega$ is an elliptic pseudodifferential operator of order $-1$.*

**Proof** Using a resolvent identity, we see that

$$(ND)_K - (ND)_\Omega = \left(B + \frac{1}{2}I\right)^{-1}\left(B - \frac{1}{2}I\right)^{-1} A.$$

It follows from the asymptotic expansion of $G_0(x, y)$ in Lemma 2.1 that $A$ is an elliptic pseudodifferential operator of order $-1$. Moreover the operator $B$ is compact, so the other two factors are pseudodifferential operators of order $0$ which do not change the principal symbol noticeably. □

Let us now return to the expression of $V$. We have

$$\langle f, Vg \rangle = \int_\Gamma \partial_N u \cdot \overline{(v_\Omega - v_K)} \, dS$$

$$= \int_\Gamma \partial_N u \cdot \overline{(v_\Omega - w + w - v_K)} \, dS$$

$$= \int_\Gamma \partial_N u \cdot \overline{\big((ND)_\Omega(\partial_N(v_\Omega - w)) + (ND)_K(\partial_N(w - v_K)))\big)} \, dS$$

$$= \int_\Gamma \partial_N u \cdot \overline{\big(((ND)_K - (ND)_\Omega)(\partial_N w)\big)} \, dS.$$

Since we are interested in $T_\mu$ and not $V$, we may assume that $f$ and $g$ belong to $\mathfrak{L}_{\Lambda_\mu}$. Then $u = Rf = \Lambda_\mu^{-1} f$ and $w = Rg = \Lambda_\mu^{-1} g$. For such $f$ and $g$ we get

$$\langle f, Vg \rangle = (\Lambda_\mu)^{-2} \int_\Gamma \partial_N f \cdot \overline{\big(((ND)_K - (ND)_\Omega)(\partial_N g)\big)} \, dS$$

or, with the introduced operators above

$$\langle f, Vg \rangle = (\Lambda_\mu)^{-2} \int_\Gamma f \cdot \overline{\big((DN)_K^*((ND)_K - (ND)_\Omega)((DN)_K g)\big)} \, dS. \tag{4.4}$$

Moreover, $(DN)_K$ is an elliptic pseudodifferential operator of order 1. This follows from the identity $A(DN)_K = B - \frac{1}{2}I$ and the fact that $A$ is an elliptic Pseudodifferential operator of order $-1$. It follows from (4.4) that $T_\mu$ is an elliptic pseudodifferential operator or order 1.

Next, we prove the inequality (3.3). Because of the projections, it is enough to show it for functions $f$ in $\mathfrak{L}_{\Lambda_\mu}$.

*The lower bound*: We prove that there exists a subspace $\tilde{\mathcal{S}} \subset \mathfrak{L}_{\Lambda_\mu}$ of finite codimension such that the lower bound in (3.3) is valid for all $f \in \tilde{\mathcal{S}}$. Since $f \in \mathfrak{L}_{\Lambda_\mu}$ we have $L_\mu f := (L - \Lambda_\mu)f = 0$ so $f$ belongs to the kernel of the second order elliptic operator $L_\mu$. Let $\varphi = f|_\Gamma$. We study the problem

$$\begin{cases} L_\mu f = 0 & \text{in } K^\circ \\ f = \varphi & \text{on } \Gamma. \end{cases} \tag{4.5}$$

Let $E(x, y)$ be the Schwarz-kernel for $L_\mu$. It is smooth away from the diagonal $x = y$. One can repeat the theory with the single and double layer potentials for $L_\mu$ and write the solution $f$ in the case it exists.

Let $B_\mu$ be the double layer operator evaluated at the boundary,

$$B_\mu \alpha(x) = \int_\Gamma \partial_{N_y} E(x, y) \alpha(y) \, dS(y), \quad x \in \Gamma.$$

The operator $B_\mu$ is compact, since the kernel $\partial_{N_y} E(x, y)$ has a weak singularity at the diagonal $x = y$. Thus there exists a subspace $\mathcal{S}_1 \subset L_2(\Gamma)$ of finite codimension such that the operator $\frac{1}{2}I + B_\mu$ is invertible on $\mathcal{S}_1$. Hence, there exists a subspace $\tilde{\mathcal{S}} \subset \mathfrak{L}_{\Lambda_\mu}$ of finite codimension where we have the representation formula

$$f(x) = \int_\Gamma \frac{\partial E(x, y)}{\partial \nu_y} \left( \left( \frac{1}{2} I + B_\mu \right)^{-1} \varphi \right)(y) \, dS(y), \quad x \in K^\circ \tag{4.6}$$

for all $f \in \tilde{\mathcal{S}}$. The inequality $\|f\|_{L_2(K_0)} \leq C \|f\|_{L_2(\Gamma)}$ follows easily from 4.6 for all such functions $f$.

Since we also have $\|f\|_{L_2(\Gamma)} \leq C \|f\|_{H^1(\Gamma)}$ the lower bound in (3.3) follows via the lower bound in (4.2).

*The upper bound*: By the upper bound in (4.2) it is enough to show the following inequalities

$$\|f\|_{L_2(\Gamma)} \|f\|_{H^1(\Gamma)} \leq C \|f\|_{H^{1/2}(K)} \|f\|_{H^{3/2}(K)} \leq C \|f\|_{H^2(K)}^2 \leq C \|f\|_{L_2(K_1)}^2.$$

However, the first inequality is just the Trace theorem, the second is the Sobolev-Rellich embedding theorem. We note that $L_\mu f = 0$, so the third inequality is a standard estimate for elliptic operators.  □

## 5 Spectrum of Toeplitz operators in a Reinhart domain

In the case when $K$ is a Reinhart domain one can strengthen part ($b$) of Lemma 3.6. Assume that $K^\circ$, the interior of $K$, is a Reinhart domain. This means that $0 \in K^\circ$ and if $z \in K^\circ$, then the set

$$\left\{ \left(w^1, \ldots, w^d\right) \mid w^j = tz^j, \ t \in \mathbb{C}, |t| < 1 \right\}$$

is a subset of $K^\circ$. If the set

$$\log|K| = \left\{ \left(y^1, \ldots, y^d\right) \mid y^j = \log|z^j|, \ z \in K^\circ \right\}$$

is convex in the usual sense, then $K^\circ$ is said to be logarithmically convex, and $K^\circ$ is a domain of holomorphy. Denote by $V_K: \mathbb{R}^d \to \mathbb{R}$ the function defined by

$$V_K(x) = \sup_{y \in \log|K|} \langle x, y \rangle.$$

We denote by $J: \mathfrak{F}_B^2 \to \widetilde{\mathcal{H}} := L_2\left(K, e^{-\frac{B}{2}|z|^2} \, dm(z)\right)$ the embedding operator. The $s$-values $s_{\hat{\kappa}}$, $\hat{\kappa} \in \mathbb{N}^d$, of $J$ coincides with the numbers

$$\left\{ \|z^{\hat{\kappa}}\|^2_{\widetilde{\mathcal{H}}} \Big/ \|z^{\hat{\kappa}}\|^2_{\mathfrak{F}_B^2} \right\}_{\hat{\kappa} \geq 0} \tag{5.1}$$

Unlike the case $d = 1$, see [FP06], it is natural to numerate the eigenvalues by the $d$-tuples $\hat{\kappa} = (\kappa_1, \ldots, \kappa_d)$, just as for the eigenvalues of the Laplace operator in the unit cube $[0, 1]^d$, where the eigenvalues are given by $(2\pi)^{-d}|\hat{\kappa}|_2^2 = (2\pi)^{-d}\left(\kappa_1^2 + \cdots + \kappa_d^2\right)$.

**Lemma 5.1** *Let $d > 1$ and $\omega = \hat{\kappa}/|\hat{\kappa}|$. Then*

$$(\hat{\kappa}! s_{\hat{\kappa}})^{1/|\hat{\kappa}|} \sim \frac{B}{2} \exp\left(2V_K(\omega)\right)\left(1 + o(1)\right), \quad \text{as } |\hat{\kappa}| \to \infty. \tag{5.2}$$

**Proof** The denominator in (5.1) is easily calculated to be

$$\|z^{\hat{\kappa}}\|^2_{\mathfrak{F}_B^2} = \left(\frac{2\pi}{B}\right)^d \left(\frac{2}{B}\right)^{|\hat{\kappa}|} \hat{\kappa}!.$$

For the numerator, we do estimations from above and below, as in [Par94]. First, note that

$$I_{\hat{\kappa}} = \|z^{\hat{\kappa}}\|^2_{\widetilde{\mathcal{H}}} = \int_{\log|K|} \exp(2\langle \hat{\kappa}, x \rangle) \, d\widetilde{m}(x),$$

where $d\widetilde{m}(x)$ is the transformed measure. It is clear that

$$I_{\hat{\kappa}} \leq \exp(2|\hat{\kappa}|V_K(\omega))m(K).$$

For the inequality in the other direction, fix $\delta > 0$. The hyperplane

$$\langle \hat{\kappa}, x \rangle = (1-\delta)V_K(\hat{\kappa})$$

cuts $\log|K|$ in two components. Let $P_\delta$ be the component for which the inequality $\langle \hat{\kappa}, x \rangle \geq (1-\delta)V_K(\hat{\kappa})$ holds. Then we have

$$I_{\hat{\kappa}} \geq \int_{P_\delta} \exp\left(2|\hat{\kappa}|(1-\delta)V_K(\omega)\right) d\widetilde{m}(x) \geq C_\delta \exp\left(2|\hat{\kappa}|(1-\delta)V_K(\omega)\right),$$

where $C_\delta = \int_{P_\delta} d\widetilde{m}(x) > 0$. It follows that

$$(\hat{\kappa}! s_{\hat{\kappa}})^{1/|\hat{\kappa}|} \leq \left(m(K)\left(\frac{B}{2\pi}\right)^d\right)^{1/|\hat{\kappa}|} \frac{B}{2} \exp\left(2V_K(\omega)\right)$$

and

$$(\hat{\kappa}! s_{\hat{\kappa}})^{1/|\hat{\kappa}|} \geq \left(C_\delta\left(\frac{B}{2\pi}\right)^d\right)^{1/|\hat{\kappa}|} \frac{B}{2} \exp\left(2(1-\delta)V_K(\omega)\right),$$

from which (5.2) follows. □

### Acknowledgements

I would like to thank my supervisor, Professor Grigori Rozenblum, for introducing me to this problem and for giving me all the support I needed.